\documentclass[reprint,amsmath,amssymb, aps,]{revtex4-1}

\usepackage{siunitx}
\usepackage{graphicx}
\usepackage{dcolumn}
\usepackage{bm}

\begin{document}

\preprint{In press Astrobiology}

\title{Detecting microbiology in the upper atmosphere: relative-velocity filtered sampling}

\author{Arjun Berera}
\email{ab@ph.ed.ac.uk}
\author{Daniel J. Brener}%
 \email{daniel.brener@ed.ac.uk}
 \affiliation{The Higgs Centre for Theoretical Physics, University of Edinburgh, UK\\
}%
\author{Charles S. Cockell}%
\email{c.s.cockell@ed.ac.uk}
 \affiliation{UK Centre of Astrobiology, University of Edinburgh, UK}

\date{\today}
\begin{abstract}
The purpose of this paper is to re-open from a practical perspective the question of the extent in altitude of the Earth's biosphere. We make a number of different suggestions for how searches for biological material could be conducted in the mesosphere and lower thermosphere, colloquially referred to as the ignoreosphere due to its lack of investigation in the meteorological community compared to other regions. Relatively recent technological advances such as CubeSats in Very Low Earth Orbit or more standard approaches such as the rocket borne MAGIC meteoric smoke particle sampler, are shown as potentially viable for sampling biological material in the ignoreosphere. The issue of contamination is discussed and a potential solution to the problem is proposed by the means of a new detector design which filters for particles based on their size and relative-velocity to the detector.\\
\\
 In Press \textit{Astrobiology} 2023. For the purpose of open access, the author has applied a Creative Commons Attribution (CC BY) licence to the Author Accepted Manuscript.
\end{abstract}

\maketitle


\section{Introduction}

The physical pervasiveness of life forms and biological structures around the Earth has been an open and much discussed question for centuries. The idea that the Earth might be a source of life material in the Solar System, and beyond, has both philosophical and practical implications. In terms of pure ecology, the maximum altitude at which life might be found is an important question in its own right as it defines the Earth systems physical extent as a habitat. An example of a practical application is in exploring the upper atmosphere as a possible transport mechanism for pathogenic microorganisms around the Earth \cite{brown2002aerial} \cite{bebber2014global}. Furthermore it has been suggested that if biological particles can be found in the thermosphere then hypervelocity space dust has sufficient momentum to facilitate the planetary escape of such particles, which leads to astrobiology implications \cite{Berera2017}. There has been no concerted research effort placed on the mesosphere and thermosphere by the astrobiology community, despite interest \cite{astrobio_smith} \cite{dassarma_antunes_dassarma_2020} \cite{microbio_review}.

The concept of the biosphere was first introduced by the geologist Eduard Suess in 1875 as the surface on the Earth where life can be found \cite{suess_1875}. This was later extended by the Russian polymath Vladimir I. Vernadsky to be any complete system which encapsulates life \cite{Vernadsky1998}. As the word implies, a biosphere has boundaries beyond which life cannot be found. Therefore it is at the boundaries of the biosphere that one finds life at the extremes. Research has established the existence of life at extreme temperatures within volcanoes and in the arctic, at extreme pressures in the deep ocean, and in high radiation environments \cite{volcanic} \cite{Clarke2013} \cite{Krisko2013}. Certain life forms have even been shown to survive in the space environment \cite{iss_bact} \cite{micro_org_space}. The extent of the biosphere in the upper atmosphere, i.e. the existence of life at extreme altitudes, remains comparatively little explored.

In certain meteorological circles, the mesosphere and lower thermosphere (MLT) is colloquially referred to as the ``ignore-osphere" because its unique position makes it inaccessible via balloons and conventional aircraft due to the low air density. Its also inaccessible for satellites as there is still sufficient air for drag to be non-negligible \cite{stern2010next}. Thus until recently, the only viable sampling option has been to use sounding rockets. In the last decade new possibilities for sampling have arrived, such as small inexpensive CubeSats and low Earth-orbit (LEO) commercial space craft \cite{cubesat_review} \cite{vleo_review}.

The highest altitudes where life seems to have been plausibly established are the top of the troposphere and the lower stratosphere \cite{dassarma_antunes_dassarma_2020} \cite{Bryan2019} \cite{microbio_review}. However, studies deeper into the stratosphere remain controversial due to questions of contamination. In a recent review Šantl-Temkiv et al. 2022 emphasise that caution is required when interpreting findings in studies where decontamination measures are not fully reported \cite{microbio_review}. Earlier studies do not fully report decontamination measures or evaluate the issue compared to modern standards. However, there is a consistent literature dating to the present which reports micrometre sized particles in the stratosphere e.g. \cite{Heintzenberg} \cite{rosen} \cite{ursem} \cite{XU2003201} \cite{yinyan} \cite{microbio_review}. Additional studies found that these particles include bacteria \cite{Griffin2004} \cite{ursem} \cite{Wainwright2003} \cite{microbio_review}. The most recent observational campaign of the stratosphere found two bacteria \textit{Bacillus simplex, Staphylococcus pasteuri} and a fungus, \textit{Engyodonitium album} at 41 km \cite{Wainwright2003}.

The purpose of this paper is to entirely re-open the question of what the upper limit of the Earth's biosphere is with practical calculations and suggestions for field campaigns. Our paper builds on the pre-existing observations of the lower atmosphere and on recent work which has identified strong vertical winds, amongst other mechanisms such as volcanic eruptions, as capable of ejecting nanometre sized particles into the MLT \cite{doi:10.1098/rspa.2021.0626}.

The highest altitude mission that sought to determine the extent of the biosphere was conducted in 1974 by Imshenetsky et al. of the Institute of Microbiology, USSR Academy of Sciences \cite{Imshenetsky1978}. They used sounding rockets fitted to capture microbiological samples up to 77 km in the mesosphere. They reported finding fungal spores. These findings are the only published study of these altitudes, and therefore there remain concerns about contamination. We propose a new approach to address these questions. Only through multiple and independently run studies will it be possible to settle concerns around contamination, as was done for determining the existence of life in the deep ocean \cite{oceans1} \cite{Escudero2018}.

Samples from the surface of the International Space Station (ISS) have been found to have DNA from several kinds of bacteria which were genetically similar to those found in the Barents and Kara seas' coastal zones \cite{ISS_DNA}. It was proposed that the wild land and marine bacteria DNA could transfer from the lower atmosphere into the ionosphere-thermosphere using the ascending branch of the global electric circuit. However, a more likely explanation would be contamination due to spacecraft passing through the lower atmosphere. Atmospheric transmission of microbes has also been postulated as one of the mechanisms behind Antarctic microbial diversity changes \cite{antarctica}.

\section{MLT sampling methodologies}
\subsection{Standard approaches}
Since the late 1940s, there have been rocket sampling missions to high altitudes \cite{encar1}. These initially involved the use of evacuated steel bottles with an altitude triggered opening and closing mechanism. Vacuum tubes are a perfectly valid method for directly sampling the upper atmosphere but they are limited in terms of temporal and volumetric capacity. 

With the advent of the Cold War, it became imperative to sample the upper atmosphere for radiative fallout during nuclear weapons tests. To capture enough of these particles, larger volumes needed to be sampled, leading to the development of the first cyrogenic samplers such as the ENCAR-1 \cite{cryo} \cite{encar1}. Prior contamination of such a sampling device is readily detectable via a Geiger-counter, and there are no large natural sources of radioactive material in the atmosphere, hence these earlier studies had little to contend with contamination wise \cite{encar1}. 

The only direct searches for biological material in the MLT have been conducted by Imshenetsky et al. using a specially designed sampler located in the nose cone of a sounding rocket \cite{Imshenetsky1978}. 

They assumed that the rocket's outer casing would experience temperatures $>$1,000 Kelvin, acting as an ascent sterilisation. While this is correct in principle, aerodynamically modelling the heating profile for the rocket is the only way to be certain. The assumption by \cite{Imshenetsky1978} that the heating was uniform and therefore the sterilisation would also be, is a flaw in their contamination strategy.

Since the missions by \cite{Imshenetsky1978} there have been no further samplings of the MLT conducted looking directly for biological particles. This lead the aerobiology community to conclude 77 km as the maximum empirically defined extent of the biosphere in altitude \cite{astrobio_smith} \cite{Horneck2010-gw}. The review of Smith 2013, criticises this conclusion as Imshenetsky 1978 neglected to give details on contamination precautions other than the aforementioned drag heating of the rocket exterior or the sealing of the samples in-flight \cite{astrobio_smith}. It is crucial that future field campaigns report clearly and transparently all procedures used, even those considered routine.

The MLT has been studied by those trying to understand meteoric smoke and the formation of noctilucent clouds. The most relevant to our work is the MAGIC meteoric smoke particle sampler \cite{Hedin744223}. Meteoric smoke particles are neutrally charged and nanometre in size, making them challenging to detect. Berera et al. 2022 hypothesised that MLT large vertical winds could potentially be strong enough to push particles of this size to around 120 km in rare events \cite{doi:10.1098/rspa.2021.0626} \cite{Berera2017}. The main difficulty with sampling particles of this size is that they tend to follow the flow around the payload, not into the detector \cite{HEDIN2007818} due to the low air density complicating the aerodynamics.

This problem of the air density is a question of whether the flow is in the continuum limit. Physically this corresponds to whether the particles are dominated by individual molecular collisions, such as in Brownian motion, or by the larger scale fluid dynamics. This is determined by the number density of the air molecules with respect to the characteristic length scale of the object in the flow. MLT flow conditions range from the transition region between continuum and non-continuum to the free molecular flow regime (non-continuum). The rarefaction of a gas interacting with an object with characteristic size L, can be described by the Knudsen number, Kn$ = \frac{\lambda}{L}$, where $\lambda$ is the mean free path of the gas. For continuum flow conditions one must have $Kn<<0.1$ and for free molecular flow Kn$>10$. 

In the MLT Kn ranges from 0.1 - 10,000, hence any direct flow measuring instrument must be capable of dealing with both transition and free molecular flow, unless separate missions are conducted per regime. Flow conditions and momentum transfer also change depending on the sizes of particle considered. In the MLT particles with $L\sim$ 10 nm interact enough with the surrounding rarefied gases that their interaction can be considered continuous \cite{acp-7-3701-2007}. Most viruses have a radii of around 20 nm, therefore if one only wanted to search for whole viruses and larger bacteria or fungi, then simple continuum modelling may be appropriate. However, accurate modelling of the effect of the gas on heating the detector would need to be carried out.

\subsection{New approach: relative-velocity filtered sampling}
Here we propose a new technique for reducing the risks of contamination effecting sampling at high altitudes. A detector needs to be designed that can sample high volumes over long periods of time in order to have a reasonable chance of encountering a biological particle. Detectors which impede the airflow are not appropriate, as this will increase drag, reducing the flight time and thence the total sampled volume per mission.

For a detector continuously sampling a flow, the particles are moving with respect to the detector rest frame. We propose that most contaminants from craft interior and exterior will likely be approximately at rest with respect to the detector. Therefore by eliminating particles which satisfy that criteria, one reduces an aspect of the contamination problem.

In terms of detector design, it is easier to apply this thinking by considering the velocities of the particles. Contaminants from the payload delivery system will have a velocity smaller than particles entering the detector as part of the flow generated by the motion of the rocket or orbiting craft. Contaminants are most likely to come from the exterior of the crafts surface and due to the low density at these altitudes,  will take longer to reach velocities close to the mean flow. 

Therefore if the detector moves sufficiently fast, $>$ an order of magnitude say, relative to the rarefied gas flow, then any particle that detaches itself from the exterior of the craft entering the detector will have a slower velocity. This could allow for such a detector to be placed at a position other than at the head of the rocket. However careful simulation and modelling would be required to ensure the correct relative velocity is selected for an accurate trigger action.

\begin{figure}[htb!]
    \centering
    \includegraphics[scale=0.5]{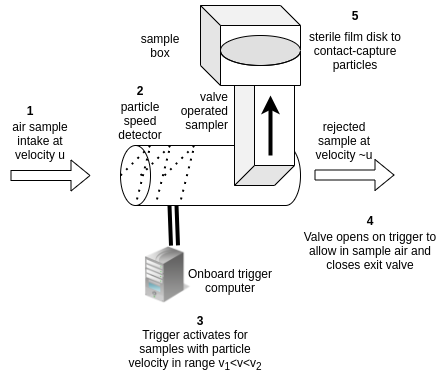}
    \caption{Trigger mechanism key stages.}
    \label{fig:trigger_sampling}
\end{figure}

In Figure \ref{fig:trigger_sampling} we have drawn a high-level schematic of the triggering process for taking a sample. Once the craft is at the selected sampling altitude, the sampling probe should open to allow the air to pass through it. After some short time, the triggering mechanism is activated. To detect a particle of interest to capture, a small lidar system could be used to scan the incoming flow in the tube detecting both particle size and velocity. The triggering computer will have two options. It can either allow the flow to continue through the detector or it can close a valve, forcing the flow to pass into a sampling chamber where particles can be captured. The decision tree is,
\begin{enumerate}
    \item IF $v<u$: valve open, particle is rejected from analysis
    \item IF $v\sim u$ or $v>u$ AND size $> 10.0$ nm: valve close, particle accepted for analysis
\end{enumerate}
where $u$ is the relative velocity between the analyser and the air flow, which would approximately be Mach 3, and $v$ is the speed of the particle detected by the lidar relative to the analyser.

Let us make an order of magnitude estimate for the scanning frequency and trigger speed requirements of such a detector. For simplicity, lets assume the flow into the detector travels at 1,000 m/s (approximately Mach 3, a typical sounding rocket velocity). This means that for a particle with characteristic size of 1 nm, a detector of length 10 cm would have $\sim$100$\mu$s to detect it, which is within the operating capabilities of circuitry.

Dyes which bind to specific proteins or nucleic acids (e.g. Sybr GREEN) \cite{Vitzthum1999-ts}, could be used to illuminate biological material. Currently these take time to bind and illuminate but in principle one could develop an instantaneously binding dye. The air could be sampled, sprayed and then accepted/rejected for analysis based on the detection of fluorescence. An alternative procedure could be to spray the samples once they have been filtered by relative-velocity, as a means of highlighting regions of interest on the film discs. A similar idea has been proposed, not for detection of life, but to eliminate contamination concerns by painting the exterior rocket surface with fluorescent beads to trace and identify contamination pathways \cite{2011ESASP.700..467J} \cite{astrobio_smith}.

There are a number of engineering challenges which would need to be dealt with. The crucial problem is that the detector must function for the transition and free molecular regimes. To do this Direct Simulation Monte Carlo (DSMC) models of the detector will be required. This technique was used for the MAGIC detector experiments as well as the development of other MLT sounding rocket instrumentation \cite{acp-7-3701-2007}. Secondly, the scanning device itself will be a challenge to engineer such that it can operate at high frequency and be relatively compact. There are a number of similar devices already developed for industry as detecting small particles in a flow is a standard problem (see e.g. \cite{Sudo:07}).

However, this technique to mitigate contamination during sampling is only as good as the weakest point in a mission: from craft assembly through to post-sampling processing. These are specific challenges in microbiology due to the wide range of biomasses which need to be analysed. This problem was tackled in \cite{Luhung2021}, where they developed a four-stage ultra-low biomass pipeline: amassment, storage, extraction and nucleic acid analysis. They use several decontamination procedures, including a negative control, where they processed a sample collected with zero airflow into their detector for 1 minute. Such a procedure should be implemented for our proposed technique. 

The problem of contamination in low microbial biomass microbiome studies has been thoroughly reviewed in Eisenhofer et al. 2019 and we strongly support the recommendations to minimise the influence of contaminant DNA in atmospheric sampling missions \cite{Eisenhofer2018-cj}. Another approach to contamination control would be to deliberately introduce artificial contaminants, such as polystyrene spheres, onto the surfaces of locations from which contamination might occur. The detection of these spheres by the capture device would allow for contamination to be quantified and identified. Similar approaches are regularly used in deep subsurface microbiological drilling e.g. \cite{10.1130/2009.2458(40)}, whereby fluorescent polystyrene spheres are added to the drilling mud to control for the internal contamination of core material. In summary, contamination can never be completely ruled out, as with other low biomass studies, but similarly to deep drilling and core recovery in low biomass environments, there are a number of ways to control for, and quantify, the potential for contamination in the instrument, which would be a critical part of instrument design and development.

\section{Mission strategies}
Firstly, we consider the use of sounding rockets, even simply to repeat the experiments by \cite{Imshenetsky1978}. Using a sounding rocket is highly beneficial as it provides a high sampling volume for relatively low cost. To zeroth order, even in the free molecular regime, the sampled volume will be given by the product of the flow rate through or over the surface of the detector with the sampling time. The flow rate to zeroth order is the surface area of the detector orthogonal to the flow product with the mean speed of the flow relative to the detector.

Assuming a relative velocity of 1,000 m/s and a circular detector area of 0.007 m$^2$, corresponding to a diameter of 10 cm, then one finds a flow rate of $\sim$7 m$^3$/s. A typical sounding rocket flight through the MLT lasts around 80 seconds, which would result in $\sim$560 m$^3$ of sampled volume. To find one particle of biological material in such a volume would be a significant result in the MLT. The key factor in the total volume sampled is the time the rocket can spend in the MLT. Even just 20 minutes of horizontal flight at a fixed altitude would yield a sample volume of $\sim$8,400 m$^3$. There are groups working on sampling rockets capable of such flights \cite{privat_comm_Boris}. Whilst these calculations are extremely crude, they nonetheless demonstrate real potential for defining the biosphere even for extremely low particle concentrations.

One could devise a series of sounding rocket missions where the rocket cruises at a fixed altitude while sampling, at intervals of 10 km from 50 - 150 km. These should be launched months apart to avoid inter-launch contamination. Geospatial separation of the launch sites would also be ideal as \cite{doi:10.1098/rspa.2021.0626} identified the poles as more likely to have biological material, since these areas experienced stronger vertical winds due to geomagnetic activity.

\begin{figure}[htb!]
    \centering
    \includegraphics[scale=0.45]{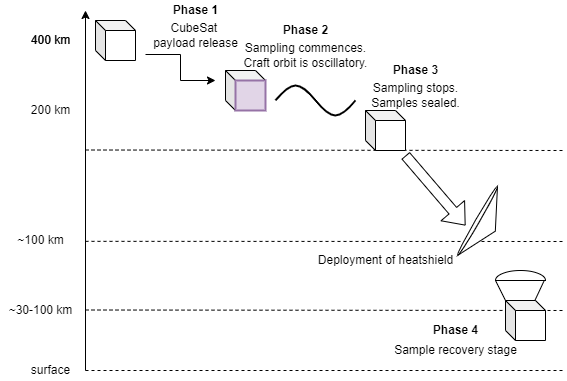}
    \caption{CubeSat mission strategy core phases.}
    \label{fig:cubesats}
\end{figure}

Of vital importance when considering the use of rockets is the aerodynamic design of the rocket and detector itself. This is to ensure that a steady, representative flow passes through the device and that when particles are selected for capture, blow-back is not an issue. Future work should examine ways to ensure that the particles captured at high speed have their momentum sufficiently arrested so that they stick to the sampler. These are crucial problems which have been around for some time in rocket-based sampling \cite{BIRD1988921} \cite{horanyi1999simulation} \cite{gumbel2001aerodynamic} \cite{HEDIN2007818}.

Alternatively, CubeSats represent a significant advance in technology in the last decade. We propose an ambitious potential mission program that would use multiple CubeSats launched months apart to sample both the upper thermosphere down to the lower thermosphere, prior to re-entry. The commercial space industry has already developed CubeSats which can orbit for short periods in what is called the Extemely Low Earth Orbit (ELEO) region, and there have already been field campaigns using multiple of such satellites to explore this region \cite{MILLAN20191466}.

One example of such a CubeSat is the TSAT Globalstar ELaNa-5 ELEO satellite (ELEO-SAT) which has dimensions 10x10x20 cm and an orbit velocity of $\sim$7 km/s \cite{tsat}. Using the same simple calculation method as before, this has the potential to sample $\sim$4,200 m$^3$ in 1 hour. Without a propulsion system, it is not possible to control the altitude of the CubeSat, and any such system would add too much additional weight which would reduce the sampling time. Hence, the CubeSat rapidly decreases in altitude and so cannot be used to sample a single altitude accurately. 

However, there are two key benefits to CubeSats over conventional sounding rockets: firstly, CubeSats can sample for hours and even weeks, whereas rockets are limited to periods minutes. Secondly, they sample the atmosphere globally, as they are in orbit. The ELEO-SAT is designed for atmospheric sampling between 120 and 325 km with over 600 orbits adding up to around 5 weeks flight time. To isolate altitude ranges, we suggest that a series of CubeSat's are deployed to sample different 100 km bands (e.g. 200 km to 100 km). 

In Figure \ref{fig:cubesats} we have illustrated the mission strategy. In phase 1, the CubeSat is launched and deployed by rocket. The CubeSat could be projected out of the rocket forwards by an explosive charge so that it remains upstream of the rocket on its first orbit ($\sim$ 90 minutes). For phase 2 the CubeSat will sample a region of atmosphere consistent with it orbital resonance as CubeSats can rise and fall by as much as 50 km in a single orbit (see e.g. Figure 16 of \cite{tsat}).

Alternatively, a single ELEO-SAT released from 400 km, travelling on average at 7 km/s for 325 km down to 120 km over 5 weeks (840 hours) would have come into contact with approximately 10$^5$ m$^3$ of air. Due to the high mixing rates at these altitudes it is extremely improbable that the satellite could sample identical volumes of air.

CubeSats usually burn up in the mesosphere at around 110 km, however it is necessary that the craft can be fully recovered. This could be achieved by using a commercially available system called a CubeSat Deorbit and Recovery System (DRS) \cite{DRS_cube}. The CubeSat DRS is a module which can be attached to the CubeSat containing a 1.2 m tension cone heat shield and parachute, all weighing $<$1.5 kg. This module is triggered to deploy the shield immediately prior to re-entry  \cite{DRS_cube}.

\section{Discussion}

It will never be possible to entirely rule out contamination from a human launched craft. We would argue that is not an issue for defining the biosphere as the increased distribution of microbiological material in the upper atmosphere and possibly into space, is now most likely a fact due to human progress. This raises many questions: What is the extent of this? Does the microbiology survive, and does it continue to breed? The astrobiology and aerobiology communities need to consider these questions carefully moving forwards.

If naturally occurring biology is to be found in the MLT it is most likely to be on the scale of viruses or smaller bacterial organelles or possibly fungal spores \cite{doi:10.1098/rspa.2021.0626} \cite{Imshenetsky1978}. If one recovers viable biological material then one could use such samples, together with the flight data, to better quantify the proportion of particles that are viable as a function of altitude. This could provide valuable data on microbial viability, survival, and destruction in the upper atmosphere.

Relative-velocity filtered sampling might also be applicable to searches for evidence of life on other planets. Conventional approaches require the landing of probes on the planet's surface, excavating the soils and performing analysis on samples \cite{Anderson_gale} \cite{life11060539}. Our proposal solves two problems: significantly reducing a major part of the contamination concerns, and removing the requirement of landing a probe in order to sample for life. Of course, a land based relative-velocity filter device would not eliminate the need for landing, although it could potentially reduce the need to move the probe around the planet. Using the natural winds to sample the biosphere is, in our opinion, likely to be a more efficient and lower risk strategy than is currently achieved with rovers. An added advantage is that being at ground level, if there is any biology on the planet, it is more likely to be found closer to the surface, assuming the planet has a similar atmospheric structure to Earth \cite{astrobio_smith} \cite{doi:10.1098/rspa.2021.0626}.

\section{Concluding remarks}

Defining the biosphere's extent in altitude is of interest to the microbiology and aerobiology communities, and has implications for ideas in astrobiology. Very few field campaigns have been conducted to empirically determine the biosphere, largely due the difficulty of sampling the MLT. As with previous microbiolgical campaigns looking for life at the extremes, the contamination problem must be foremost in any effort to sampling these remote atmospheric regions. We propose that a sampling device could detect the relative-velocity and size of incoming particulates, and thereby filtering sampled air to mitigate against self-contamination. 

The main underlying feature of the relative velocity filtered sampling method we propose, is that the particle of interest that is to be measured, should be at a different relative velocity to the detector, thus eliminating contamination that may persist on the detector. The proposal in our paper is to utilize this method to look for biological particles in the upper atmosphere, above 100km, which would be a novel discovery. The detector should be on a moving craft, such as as a sounding rocket, satellite, or even a drone, which intakes the surrounding air at high velocity and then detects the particles of interest in flight.

We have investigated how such a device could be designed around a CubeSat craft in very low Earth orbit or in the nose cone of a rocket, to enable sampling of the MLT. The ideas involve several engineering challenges, specifically around the rapid detection of particles using a lidar that is light-weight, small and of sufficient temporal and spatial resolution. Other technologies that we are unaware of may be more appropriate or in the future be developed, but the principle of using the relative-velocity of individual particles in a flow will in general always be applicable.
\section*{Acknowledgment}
The authors were supported by the United Kingdom Science and Technology Facilities Council, with Charles S. Cockell under grant ST/V000586/1.

\bibliographystyle{unsrt}
\bibliography{apssamp}

\end{document}